\documentclass[conference]{IEEEtran}

\usepackage{epigraph}
\usepackage{amssymb}

\epigraphsize{\small}
\begin{document}

\title{On Formalisms for Dynamic Reconfiguration of Dependable Systems}

\author{
\IEEEauthorblockN{Anirban Bhattacharyya, Andrey Mokhov, Ken Pierce}
\IEEEauthorblockA{Newcastle University, UK} \and
\IEEEauthorblockN{Manuel Mazzara}
\IEEEauthorblockA{Politecnico di Milano, Italy}
}

\maketitle

\begin{abstract}
\noindent
Three formalisms of different kinds -- VDM, Maude, and basic ${\bf CCS^{dp}}$ --
are evaluated for their suitability for the modelling and verification of dynamic software reconfiguration
using as a case study the dynamic reconfiguration of a simple office workflow for order processing.
The research is ongoing, and initial results are reported.
\end{abstract}

\IEEEpeerreviewmaketitle

\section{State of the Art}\label{sec:related}

We define the dynamic software reconfiguration of a system to be the change at runtime of the system's software components
or their communication links, or the hardware location of the software components or their communication links.
Dynamic software reconfiguration can help to increase the reliability of a system, and thereby increase its dependability.
However, predicting the behaviour of the system during its dynamic reconfiguration is still a challenge \cite{kn:Bha13},
to which our research is a response. Existing research in dynamic software reconfiguration can be grouped into three cases:

\begin{enumerate}
\item The reconfiguration is near-instantaneous and its duration is negligible in 
comparison to the durations of application tasks. Executing tasks in the old configuration that 
are not in the new configuration are aborted. This is the traditional method of software 
reconfiguration, and it is applicable to small, simple systems running on a uniprocessor.
	
\item The duration of the reconfiguration interval is significant in comparison to the 
durations of application tasks, and any executing task in the old configuration that can interfere 
with a reconfiguration task is either aborted or suspended until the reconfiguration is complete.
This is the most common method of software reconfiguration (see \cite{kn:SteVolKho97}, 
\cite{kn:AlmEtAl01}, \cite{kn:BloDay93}, and \cite{kn:KraMag90}), and it is applicable to some large, 
complex, distributed systems.	

\item The duration of the reconfiguration interval is significant in comparison to the durations of 
application tasks, and tasks in the old configuration execute concurrently with reconfiguration 
tasks. This method avoids aborting tasks and reduces the delay on the application due 
to reconfiguration, but it introduces the possibility of functional and temporal interference 
between application and reconfiguration tasks.
If the interference can be controlled, 
then this method is the most suitable for large, complex, distributed systems, including 
hard real-time systems, but it is also the least researched method of software 
reconfiguration. 
Existing research has focused on temporal interference between application and reconfiguration 
tasks, and on achieving schedulability guarantees 
(see \cite{kn:ShaRajLehRam89}, \cite{kn:TinBurWel92}, \cite{kn:Foh94}, \cite{kn:Ped99},
\cite{kn:Mon04}, \cite{kn:FisWin05}, and \cite{kn:Far06}).
There is little research on formal verification of functional correctness in the presence of functional interference
between application and reconfiguration tasks (see \cite{kn:MedTay00} and \cite{kn:BraEtAl04}).
\end{enumerate}

\section{Case study}\label{sec:casestudy}

We use as a case study the dynamic reconfiguration of a simple office workflow for order processing,
commonly found in large and medium-sized organizations \cite{kn:EllKedRoz95}.
These workflows typically handle large numbers of orders.
Furthermore,  the organizational environment of a workflow can change in structure, procedures, 
policies, and legal obligations in a manner unforeseen by the original designers of the workflow.
Therefore, it is necessary to support the unplanned change of these workflows.
Furthermore, the state of an order in the old configuration may not correspond to any state of the order
in the new configuration. These factors, taken in combination, imply that instantaneous 
reconfiguration of a workflow is not always possible; neither is it practical to 
delay or abort large numbers of orders because the workflow is being reconfigured. 
The only other possibility is to allow overlapping modes for the workflow during its reconfiguration.

The workflow consists of a graph of activities and the configuration is the structure of this graph.
Initially, the workflow executes in Configuration 1 and must satisfy the requirements on Configuration 1.
Subsequently, the workflow must be reconfigured 
through a process to Configuration 2 in such a way that the requirements on the reconfiguration 
process and on Configuration 2 are satisfied. The full details of the two configurations 
and their respective requirements are given in \cite{kn:MazEtAl11}.

In order to achieve a smooth transition from Configuration 1 to Configuration 2 of the workflow,
the process of reconfiguration must meet the following requirements:

\begin{enumerate}
   \item Reconfiguration of a workflow should not necessarily result in the rejection of an order.
   \item Any order being processed that was accepted \textbf{before} the start of the reconfiguration must satisfy
         all the requirements on Configuration 1.
   \item Any order accepted \textbf{after} the start of the reconfiguration must satisfy all the requirements on Configuration 2.
   \item The reconfiguration process must terminate.
\end{enumerate}

\section{Comparison of Formalisms}\label{sec:comparison}

As discussed in previous work (see \cite{kn:MazBha10} and \cite{kn:AboEtAl12}),
research shows that no single existing formalism is ideal for representing dynamic reconfiguration \cite{kn:Wer99}.
Therefore, we are using a portfolio of three formalisms of different kinds --
$VDM$ (based on the model-oriented approach), $Maude$ (based on the rewriting logic approach),
and basic $CCS^{dp}$ (based on the behavioural approach) --
to produce representations of the case study,
and to evaluate how well the different representations can be verified formally
to determine whether or not they meet the termination requirement.
We have deliberately chosen \textbf{not} to use workflow-specific formalisms
for two reasons.
First, because of our lack of fluency in them;
and second, because we believe the models should be produced using general purpose formalisms.

\section{Conclusions}\label{sec:conclusions}

Our research is ongoing. Initial results demonstrate that in all three formalisms:
functional interference between concurrently executing tasks is expressed in terms of interleaved actions, but not preemptive actions;
unplanned dynamic reconfiguration can be represented; and the termination requirement can be verified.
Basic $CCS^{dp}$ expresses reconfiguration declaratively, and (therefore) very simply,
which is useful for experimenting with different reconfiguration paths between configurations;
but there is no support for how reconfiguration can be implemented. Verification is possible through model checking.
In contrast, $VDM$ expresses reconfiguration imperatively, and (therefore) in terms of mechanisms,
which is useful for implementing reconfiguration;
but complicates experimenting with different reconfiguration paths by including unnecessary detail.
In contrast, $Maude$ expresses actions as rewrite rules, and reconfiguration as rewrite rules of rewrite rules,
which is elegant; but it is not how a typical system designer thinks about reconfiguration.


\subsubsection*{Acknowledgements}

The authors acknowledge the help given by numerous colleagues, in particular:
Jeremy Bryans, John Fitzgerald, Regina Frei, Kohei Honda, Cliff Jones, Maciej Koutny, 
Richard Payne, Traian Florin Serbanuta, Giovanna Di Marzo Serugendo and Chris Woodford.

\bibliographystyle{abbrv}
\bibliography{FA04_Bhattacharyya_NU}

\end{document}